\documentclass[twocolumn,showpacs,superscriptaddress, preprintnumbers , amsmath,amssymb,letterpaper]{revtex4}

\RequirePackage{lineno}

\usepackage{dcolumn}
\usepackage{bm}
\usepackage{amsmath}
\usepackage{multirow}
\usepackage{graphicx}
\usepackage{float}
\usepackage[colorlinks=true]{hyperref}
\begin{document}


 \newcommand{\re}{\mathop{\mathrm{Re}}}
 \newcommand{\im}{\mathop{\mathrm{Im}}}
 \newcommand{\D}{\mathop{\mathrm{d}}}
 \newcommand{\I}{\mathop{\mathrm{i}}}
 \newcommand{\E}{\mathop{\mathrm{e}}}
 \newcommand{\unite}[2]{\mbox{$#1\,{\rm #2}$}}
 \newcommand{\myvec}[1]{\mbox{$\overrightarrow{#1}$}}
 \newcommand{\mynor}[1]{\mbox{$\widehat{#1}$}}
 \newcommand{\rmsemit}{\mbox{$\tilde{\varepsilon}$}}
 \newcommand{\mean}[1]{\mbox{$\langle{#1}\rangle$}}


\date{\today}
\title{Formation and Acceleration of Uniformly-Filled Ellipsoidal Electron Bunches Obtained via Space-Charge-Driven Expansion from a Cesium-Telluride  Photocathode}

\author{P. Piot} \affiliation{Northern Illinois Center for
Accelerator \& Detector Development and Department of Physics,
Northern Illinois University, DeKalb IL 60115,
USA} \affiliation{Accelerator Physics Center, Fermi National
Accelerator Laboratory, Batavia, IL 60510, USA}
\author{Y.-E Sun}  \affiliation{Accelerator Physics Center, Fermi National
Accelerator Laboratory, Batavia, IL 60510, USA}
\author{T. J. Maxwell\footnote{current address: Stanford Linear Accelerator Center, Menlo Park, CA 94025, USA}} \affiliation{Northern Illinois Center for
Accelerator \& Detector Development and Department of Physics,
Northern Illinois University, DeKalb IL 60115,
USA} \affiliation{Accelerator Physics Center, Fermi National
Accelerator Laboratory, Batavia, IL 60510, USA}
\author{J. Ruan}  \affiliation{Accelerator Division, Fermi National
Accelerator Laboratory, Batavia, IL 60510, USA}
\author{E. Secchi\footnote{permanent address: Politecnico di Milano,  20156 Milano, Italy}}  \affiliation{Accelerator Physics Center, Fermi National
Accelerator Laboratory, Batavia, IL 60510, USA}
\author{J. C. T. Thangaraj}  \affiliation{Accelerator Physics Center, Fermi National
Accelerator Laboratory, Batavia, IL 60510, USA}

\begin{abstract}
We report the experimental generation, acceleration and characterization of a uniformly-filled electron bunch obtained via space-charge-driven expansion (often referred to as  ``blow-out regime'') in an L-band  (1.3-GHz) radiofrequency photoinjector. The beam is photoemitted from a Cesium-Telluride semiconductor photocathode using a short ($<200$~fs) ultraviolet laser pulse. The produced electron bunches are characterized with conventional diagnostics and the signatures of their ellipsoidal character is observed. We especially demonstrate the production of ellipsoidal bunches with charges up to $\sim0.5$~nC corresponding to a $\sim20$-fold increase compared to previous experiments with metallic photocathodes. 
\end{abstract}
\pacs{ 29.27.-a, 41.85.-p,  41.75.Fr}

\maketitle

\section{Introduction}
Three-dimensional uniformly-filled ellipsoidal charge distributions produce space charge fields that have a linear dependence on
position within the distribution~\cite{Kellogg,KV}. The resulting density distributions are therefore immune to space-charge-induced phase space dilution. Besides mitigating emittance dilution, these ellipsoidal bunches are notably less prone to halo formation thereby making these distributions attractive for, e.g., high-average-power free-electron lasers (FELs)~\cite{neil}. 

A scheme to generate such distributions using photo-emission electron sources was proposed by Serafini~\cite{serafini} and subsequently refined by Luiten {\em et al.}~\cite{luiten}. The latter proposal uses an ultrashort laser impinging on a prompt photo-emitter subject to a strong accelerating electric field $E_0$. The operating parameters of the electron source are chosen such that the distribution evolution is dominated by linear space charge force.
This space-charge-dominated expansion, also referred to as the ``blow-out regime''~\cite{serafini}, is achieved provided the condition~\cite{luiten}
\begin{eqnarray} \label {eq:blowout}
\frac{eE_0c\tau_l}{mc^2} \ll \frac{\sigma_0}{\epsilon_0 E_0} \ll 1
\end{eqnarray}
is fulfilled. Here $\tau_l$, $c$, $\epsilon_0$, $m$ and $e$ are respectively the duration of the photoemission process, the speed of light, the electric permittivity of vacuum, and the electronic mass and charge. The parameter $\sigma_0 \equiv Q/(\pi r^2)$ is the initial surface charge density ($Q$ is the bunch charge and $r$ the radius of the laser transverse distribution on the photocathode). 

\begin{figure}[hhhhh!!!!!!!!!!!!]
\centering
\includegraphics[width=0.49\textwidth]{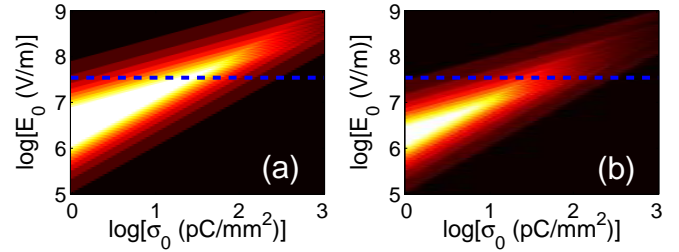}
\caption{Domain of existence of the blow-out regime (lighter colors) in the $(\sigma_0, E_0)$ parameter space for $\tau_l=50$~fs (a) and $\tau_l=200$~fs (b). The horizontal blue dashed lines correspond to $E_0=35$~MV/m (Figure adapted from Ref.~\cite{Oudheusden}).}\label{fig:domain}
\end{figure}

Considering a L-band radiofrequency (rf) gun with a typical peak accelerating field of $E_0=35$~MV/m at the photocathode surface, the conditions for ellipsoidal bunch generation using the blow-out regime can in principle be realized with $\tau_l=50$~fs; see Fig.~\ref{fig:domain} (a). A longer pulse length $\tau_l =200$~fs can still support the scheme; see Fig.~\ref{fig:domain} (b). Further increasing  $\tau_l$ would require  lower values of $E_0$ and $\sigma_0$  to meet the blow-out regime conditions (Eq.~\ref{eq:blowout})  resulting in operating points that might not be relevant to applications demanding significant (sub-nC) charge per bunch, e.g., high-average-power FELs~\cite{dc,fernando}.

For prompt-emission photocathodes, $\tau_l$ is comparable to the laser pulse duration and the formation of ellipsoidal bunches was
experimentally confirmed~\cite{pietro,moody,Oudheusden}. For semiconductor high-quantum-efficiency photocathodes, the slower photoemission response might affect the production of ellipsoidal bunches via the blow-out regime as pointed out in Ref.~\cite{JRBNIM}. 

To date, there has been no conclusive investigation regarding the  compatibility of the blow-out regime with semiconductor photocathodes ~\cite{oshea,comment}. Such an investigation is the prime purpose of this paper. In addition, the presented experimental results are obtained in a L-band photoinjector with $E_0\simeq 35$~MV/m whereas previous successful experiments have been carried out at S-band photoinjectors with higher values of typically $E_0 \sim  100$~MV/m~\cite{Oudheusden,pietro}.  The present work supports the use of the blow-out regime in low-frequency electron guns with limited peak field. Such guns are foreseen as sources for high-frequency multi-user FEL facilities~\cite{wifel,joe}.  
\section{The A0 photoinjector setup}
\begin{figure}[hhhhh!!!!!!!!!!!!]
\centering
\includegraphics[width=0.48\textwidth]{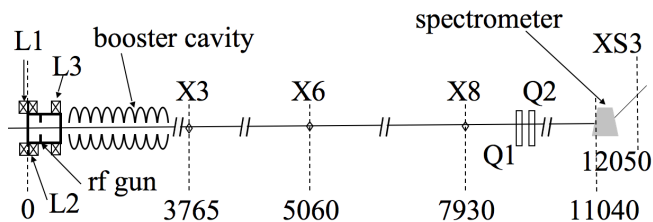}
\caption{Top view of the A0PI setup displaying elements pertinent to the present simulations and experiments. The ``L" refers to solenoidal lenses, ``X" to diagnostic stations (beam viewers and/or multi-slit masks location), and ``Q" to quadrupole magnets. The distances are in mm and referenced to the photocathode surface.}\label{fig:beamline}
\end{figure}

The numerical and experimental investigations of ellipsoidal-bunch production from a semiconductor photocathode and their subsequent acceleration to $\sim 16$~MeV  were carried out at the now-decommissioned Fermilab's A0 photoinjector; see Fig.~\ref{fig:beamline}~\cite{carneiro}.
In brief, electron bunches are generated via photoemission from a cesium telluride (Cs$_2$Te) photocathode located on the back plate
of a 1+1/2 cell radio-frequency (rf) cavity operating at 1.3~GHz (the ``rf gun"). The rf gun is surrounded by three solenoidal lenses that control the beam's transverse size and emittance. The beam is then accelerated in a 1.3~GHz superconducting rf cavity~\cite{aunes}. Downstream of the booster cavity, the beamline includes quadrupoles and steering dipole magnets, and diagnostics stations. The transverse density diagnostics are based on Cerium-dopped Yttrium Aluminum Garnet (Ce:YAG) screens (labeled as ``X" in Fig.~\ref{fig:beamline}). A multi-slit mask insertable at X3 can be used to measure the transverse emittance by analyzing the resulting beamlet's distribution at the X6 screen~\cite{lumpkin}. At its end, the beamline incorporates a horizontally-bending spectrometer equipped with a Ce:YAG screen (XS3) for
energy measurement. The horizontal dispersion value at XS3 location is $|\eta| =317$~mm. The nominal operating parameters of the A0PI
subsystems are gathered in Table~\ref{tab:param}.

\begin{figure}[hhhhh!!!!!!!!!!!!]
\centering
\includegraphics[width=0.49\textwidth]{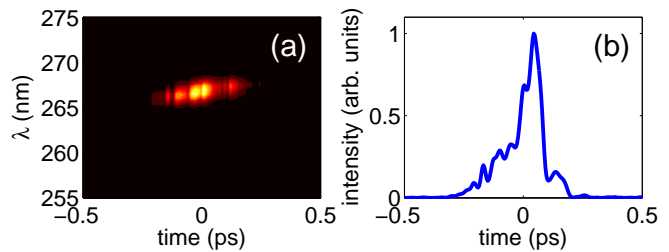}
\caption{Measured spectrogram (a) using  a polarization-gate implementation of the frequency-resolved optical gating (FROG) technique and reconstructed temporal distribution (b) of the UV pulse. The inferred rms UV pulse duration is 122 fs.}\label{fig:frog}
\end{figure}

Nominally, the A0PI's photocathode laser consists of a frequency-quadrupled Neodymium-doped Yttrium Lithium Fluoride (Nd:YLF) laser~\cite{jianliang}. Due to its narrow bandwidth ($\Delta\lambda \sim 1$~\AA), this laser system cannot  produce laser pulses with duration $< 3$~ps. Therefore a short-pulse laser based on a Titanium-Sapphire oscillator and regenerative amplifier was recently installed~\cite{timNIMA}. The system also includes an acousto-optic programmable dispersive filter  system to control the laser shape~\cite{tournois,fastlite}. The generated 3-mJ infrared (IR) pulses ($\lambda=800$~nm) with duration of $\sim 50$~fs (rms) are frequency-tripled using a two-stage frequency upconversion scheme~\cite{trippler}. The second-harmonic-generation (SHG) and sum-frequency-generation (SFG) stages  consist of $\beta$--barium borate ($\beta$-BBO) crystals of  respective thicknesses 300 and 150~$\mu$m. The frequency upconversion  was optimized to preserve the short pulse duration: the thicknesses of the $\beta$-BBO crystals were selected based on numerical simulations using a customized version of {\sc snlo}~\cite{snlo}, and a calcite crystal was included to compensate for delay accumulation as the 800-nm and SHG-generated 400-nm pulses co-propagate in the 800-nm half-wave plate needed to match the polarizations prior to the SFG stage. The resulting ultraviolet (UV) pulses have a root-mean-square (rms) duration of $\sim 150$~fs as measured using a polarization-gate version~\cite{pg1,pg2} of  a frequency-resolved optical gating (FROG) method~\cite{frog,froguv}; see Fig.~\ref{fig:frog}.  The estimated duration of the UV pulse is below $\sim200$~fs at the photocathode (after transport through a $\sim 20$-m long optical beamline and passage through three vacuum windows).

\section{Numerical Simulations}
To confirm whether the A0PI  facility could support the production of ellipsoidal bunches using the blow-out regime, particle-in-cell
simulations were performed using the program {\sc astra}~\cite{astra}  which includes a quasi-static space-charge algorithm.

The initial conditions for the electron beam generation are dictated by the
photocathode drive-laser. For the present simulation, the UV laser pulse was taken to follow a Gaussian distribution with 200-fs (rms) duration.
However,  contrary to metallic photocathodes, their semiconductor counterparts have a finite emission time generally described as a
diffusion process. Measurement performed for NEA cathode have validated this model and response time in the ps regime were
reported~\cite{hartmann,bazarov}. To our knowledge there is no corresponding experimental data for Cs$_2$Te cathodes and to date the response time were only investigated numerically via Monte-Carlo simulations~\cite{cstepaper} of the Òthree-stepÓ model~\cite{Spicer}.
\begin{table}[h!]
\caption{\label{tab:param} Nominal settings for the rf-gun,
booster cavity, and the photocathode UV laser.}
\begin{center}
\begin{tabular}{l c c}
\hline \hline parameter                       &      value       &
units  \\ \hline
laser injection phase$^a$           &  45 $\pm$ 5     & rf deg \\
laser radius on cathode         & [0.3,2]    & mm     \\
laser rms pulse duration            &  $<200$  & fs     \\
bunch charge                   &  [100$\sim$700]    & pC \\
$E_z$ on cathode                &  33.7  $\pm$ 0.2  & MV/m     \\
peak $B_z$$^b$ (L2, L3)       &  (0.158$\sim$ 0.041)   & T     \\
booster cavity acc. field    &  $\sim$ 12.0    & MV/m     \\
booster cavity phase$^c$    &  [-60$\sim$60]   & rf deg     \\
\hline \hline
\end{tabular}
\end{center}
$^a$ {\small the phase is referenced w.r.t the zero-crossing phase,}\\
$^b$  {\small the peak field of solenoid L1 was tuned to zero the axial magnetic field on the photocathode,} \\
$^c$  {\small the phase is referenced w.r.t. the maximum energy. }
\end{table}
\begin{figure}[hhhhh!!!!!!!!!!!!]
\centering
\includegraphics[width=0.49\textwidth]{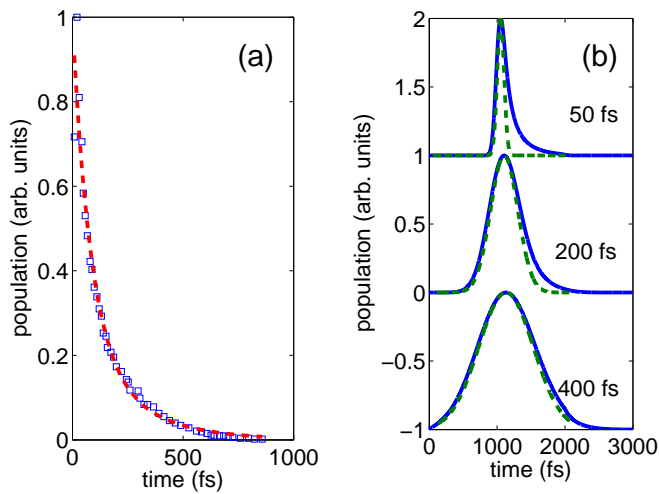}
\caption{Monte-Carlo simulation (squares) taken from Ref.~\cite{cstepaper} and parametrization (dashed line) of the electron transit time from an initial Dirac-like laser excitation (a). Impact of the slow response time on the photoemission profile (blue traces) for three different Gaussian-like laser profiles (green dashed traces) of indicated rms duration (b). The traces in plot (b) are vertically offset for clarity.  Larger time ordinates correspond to the bunch tail. }\label{fig:response}
\end{figure}
These numerical results indicate a significant spread in electron transit time for an initial Dirac-like laser excitation~\cite{cstepaper}; see
Fig.~\ref{fig:response} (a). In particular it is found that 90\% of the electrons are emitted within a temporal window of 370 fs.  The distribution also has a trailing electron population that extents to $\sim 1$~ps after the laser impinged the photocathode. In order to take into account the finite emission time in our simulation we parametrized the data shown in Fig.~\ref{fig:response} (a) with the function
\begin{eqnarray}
\Lambda(t)=a_1e^{-t/\tau_1} + a_2e^{-t/\tau_2}, 
\end{eqnarray}
 where $a_i$ and $\tau_i$ are fitting parameters. Note that the motivation of this mathematical parametrization is to provide an analytic function for simulation purpose rather than to attempt a description of the physics associated to the response profile. The temporal charge distribution during the photoemission process (at the photocathode surface) is then taken as the convolution 
 \begin{eqnarray}\label{eq:profile}
 Q(t)=\int_{-\infty}^{+\infty} Q_0(t') \Lambda(t-t') dt',
 \end{eqnarray}
  where $Q_0(t)$ is a Gaussian charge distribution with rms duration given by the drive-laser pulse duration $\sigma_t$. The results of such a convolution appear in Fig.~\ref{fig:response} (b) for different laser profiles. The latter Figure shows that for laser pulses with $\sigma_t \le 400$~fs, the finite emission time of Cs$_2$Te significantly alters the charge emission profile and especially leads to asymmetric emission profiles with long trailing tails.
\begin{figure}[hhhhh!!!!!!!!!!!!]
\centering
\includegraphics[width=0.48\textwidth]{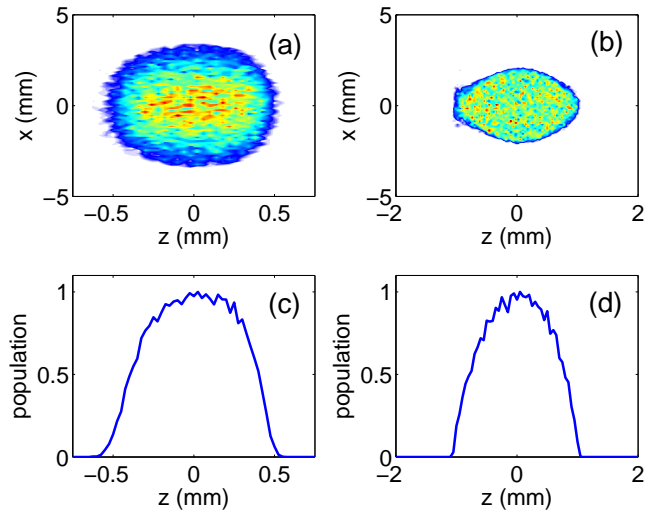}
\caption{Simulated spatio-temporal distribution $(z,x)$ at $s=0.47$~m (a) and $s=3.77$~m (b) (corresponding to X3 screen) wrt to the photocathode and associated longitudinal charge distributions [(c) and (d) respectively]. The charge is  $250$~pC and the rms laser transverse size is $\sigma_c=1$~mm.  In these plots the tail of the bunch is at $z>0$. \label{fig:evol}}
\end{figure}
In order to explore the effects associated to the finite response time of the Cs$_2$Te cathodes, the photoemission process was simulated using the emission profile described by Eq.~\ref{eq:profile}. The transverse distribution of the laser was taken as a 2$\sigma$-clipped Gaussian distribution.\\

\begin{figure}[hhhhh!!!!!!!!!!!!]
\centering
\includegraphics[width=0.48\textwidth]{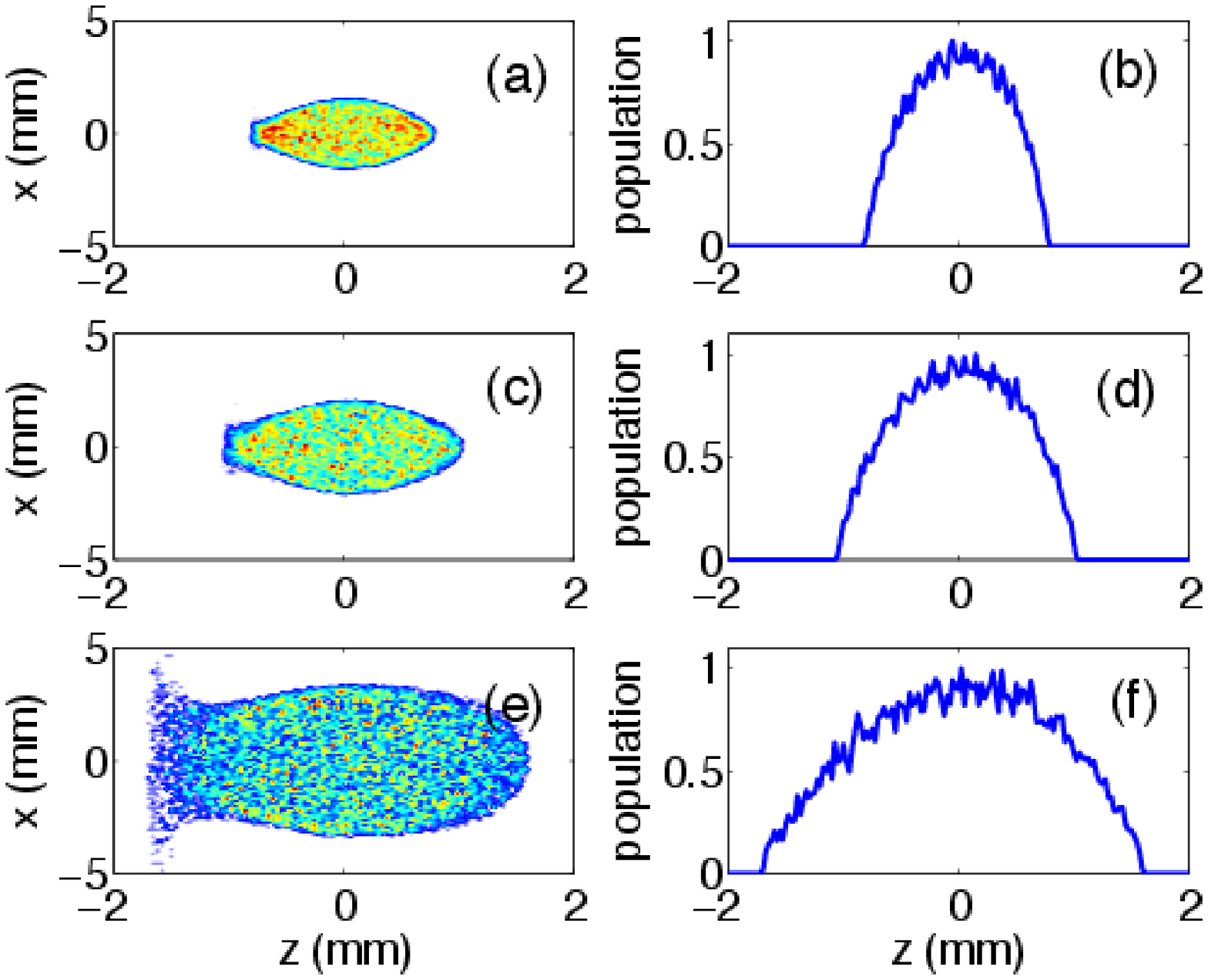}
\includegraphics[width=0.48\textwidth]{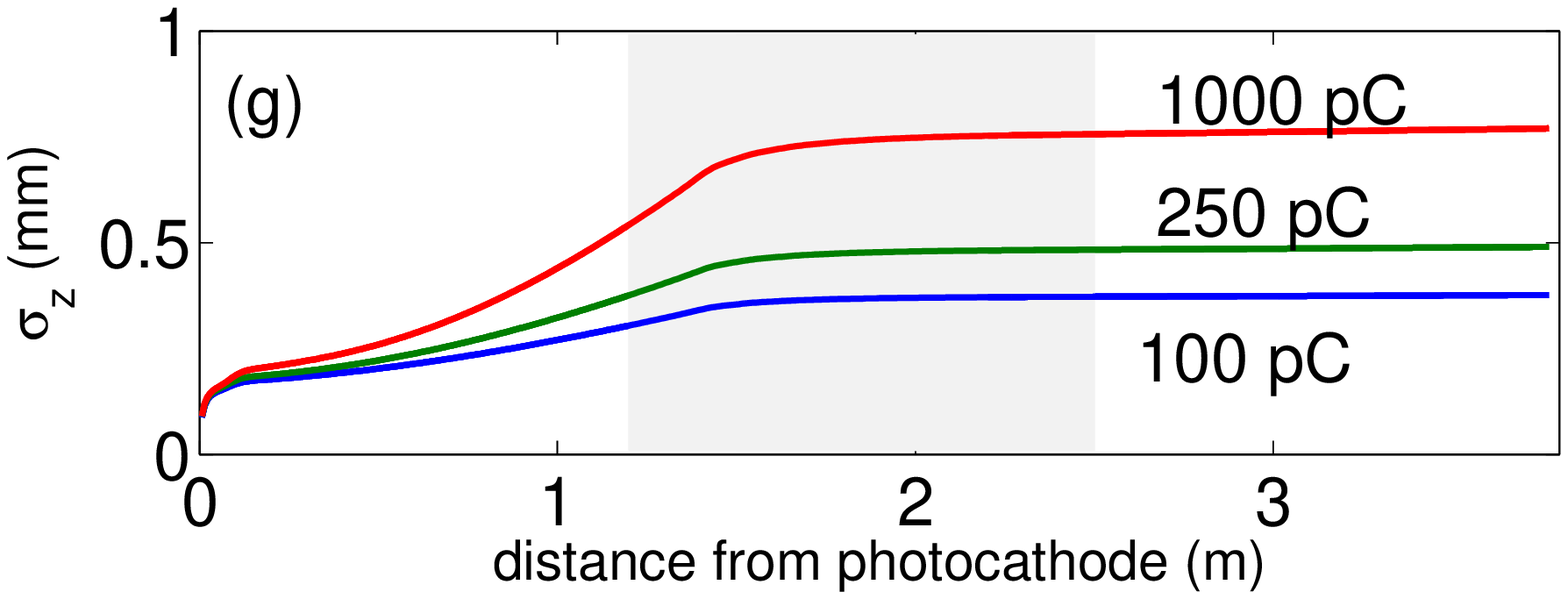}
\caption{Simulated spatio-temporal distribution $(z,x)$ (left column) and associated longitudinal charge distribution (right column) simulated at X3 ($s=3.77$~m) for three cases of charges 100 [(a), (b)], 250 [(c), (d)], and 1000~pC [(e), (f)].  In these plots the tail of the bunch is at $z>0$.  Plot (g) illustrates the rms bunch length evolution along the beamline up to $z=3.77$~m (location of X3) for the three cases of bunch charge.  The shaded area in plot (g) indicate the location of the booster cavity.}\label{fig:vsQ}
\end{figure}

 An example of simulated spatio-temporal $(z,x)$ densities and associated longitudinal charge distributions appear in Fig.~\ref{fig:evol} for a charge of $Q=250$~pC. The simulations confirm that for low-charge bunches, the A0PI can operate in the blow-out regime. Increasing the charge to higher values (up to 1000~pC) does not significantly alter the ellipsoidal character; see Fig.~\ref{fig:vsQ}. For  the range of charge attainable at the A0PI, the space-charge-driven bunch length expansion is strongly suppressed once the bunch has been accelerated by the booster cavity; see Fig.~\ref{fig:vsQ} plot (g). 

To investigate the domain of existence of the blow-out regime in the $(\sigma_0, E_0)$ parameter space using the realistic beamline and external fields of the A0 photoinjector, we introduce a figure of merit to quantify the ellipsoidal character of the bunch distribution. Since one of the properties associated to such a distribution is the linearity of the two-dimensional sub phase spaces, we introduce the mean Euclidean distance associated to the $(u,u')$ trace space  as
\begin{eqnarray}\label{eq:metric}
{\cal D}_u\equiv \frac{1}{N} \sum_{i=1}^N d_{u,i},  \mbox{~with~} d_{u,i}\equiv \frac{|u'_i-m_u u_i - q_u|}{\sqrt{1+m^2}} , 
\end{eqnarray}
where $u \in[x,y,z]$ and the summation is performed over the number of macroparticles representing the bunch, $(u_i,u'_i)$ are the trace-space coordinates of the $i^{th}$ macroparticle, and $m_u$ and $q_u$ are respectively the slope and intersect of the linear regression of the macroparticle distribution in $(u,u')$. 
In Ref.~\cite{pietro}, a different figure of merit based on the temporal asymmetry was introduced but is not as convenient to automatically compute compared to the ${\cal D}_u$. Nonetheless we verified that the two metrics were consistent. 
\begin{figure}[hhhhh!!!!!!!!!!!!]
\centering
\includegraphics[width=0.485\textwidth]{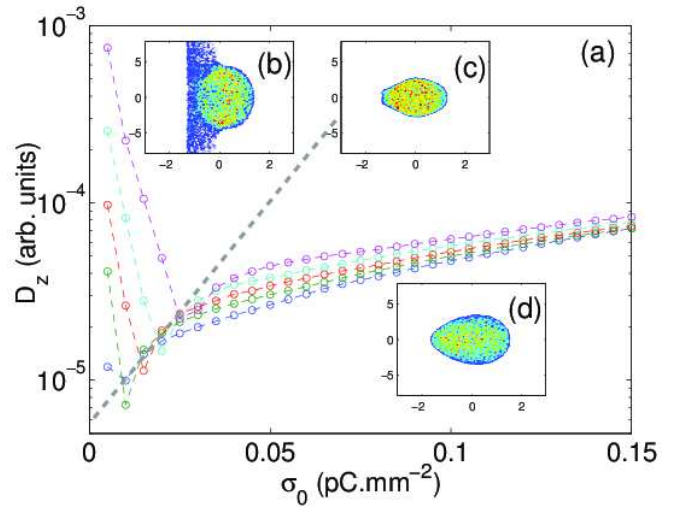}
\caption{Computed ${\cal D}_z$ metric [see Eq.~\ref{eq:metric}] (a)  for bunch charges of 100, 250, 500, 1000, and 2000 pC (shown respectively as  blue, green, red, cyan, and magenta symbols) as function of cathode surface density $\sigma_0$. The gray line indicates loci of minimum ${\cal D}_z$ and the insets (b), (c) and (d) show typical spatio-temporal distributions simulated for the three regimes delineated by the gray line; see text for details. The simulation results are evaluated at the location of X3 in Fig.~\ref{fig:beamline}. The inset plots horizontal (resp. vertical) axis corresponds to $z$ (resp. $z$) both in units of  mm. \label{fig:metric}}
\end{figure}
For the nominal gun field $E_0=35$~MV/m, the charge and transverse size of the laser were varied. For all the simulations, the photoemission temporal profile shown in Fig.~\ref{fig:response} [(b), ``200 fs" case] was used.  The results, summarized in Fig.~\ref{fig:metric}, confirm that despite the low-field L-band rf gun and the long emission time, the configuration can operate in the blow-out regime over a large range of charge. The metric ${\cal D}_z$ was chosen as it quantifies the linearity of the longitudinal phase space. We find that for each charge ${\cal D}_z$ is minimized for a slightly different value of $\sigma_0$. The locus of minima corresponds to the condition for forming an optimum ellipsoidal bunch [with spatio-temporal distribution shown in Fig.~\ref{fig:metric} (c)]. For larger values of $\sigma_0$, the image charge prevails and confers the observed  asymmetry egg-shaped spatio-temporal distribution [exemplified in Fig.~\ref{fig:metric} (d)]. Finally for low $\sigma_0$ the required beam size becomes large and leads to the alteration of the ellipsoidal character due to nonlinearity of the off-axis fields and non-linear correlation between the transverse coordinates and time [as investigated in Ref.~\cite{bass} and shown in Fig.~\ref{fig:metric} (b)].   We also note that ${\cal D}_z$ increases with the bunch charge so that low charges are more favorable to the blow-out regime for our gun configuration -- this is where most of the experimental work as been carried to date~\cite{pietro}. 
\section{Experimental results and Analysis }
Several experiments were performed to explore the operation of the A0PI in the blow-out regime. These included the measurements of the spatio-temporal $(z \propto t,y)$ distributions and associated current distribution, the estimate of LPS chirp. Finally we also measured the transverse emittances.  When possible the measurements were performed for different charge and laser spot size on the photocathode. 

\subsection{Transverse beam density}
\begin{figure}[hhhhh!!!!!!!!!!!!]
\centering
\includegraphics[width=0.485\textwidth]{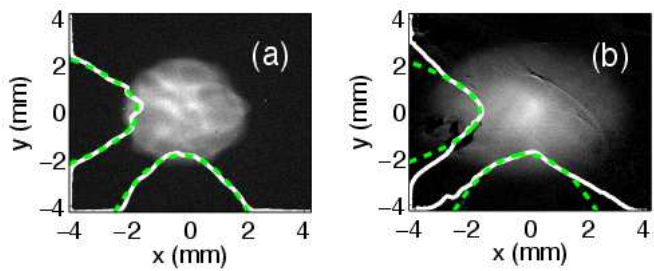}
\caption{Examples of measured  spatial $(x,y)$ distribution and associated projections (white solid lines) with parabolic fit (green dash lines). Images (a) and (b) respectively correspond to the beam being photo-emitted using the ultra-short Ti:Sp and the picosecond Nd:YLF laser systems. The beam density is measured at the Ce:YAG screen X6 (see Fig.~\ref{fig:beamline}) for $Q \simeq 250 \pm 50$~pC. }\label{fig:tden}
\end{figure}

Measuring the transverse beam density provided a first hint of the ellipsoidal character of the bunch distribution as depicted in Fig.~\ref{fig:tden} (a). The observed transverse distribution has projections that follows a parabolic distribution and comprise sharp edge as expected. Slight deviations from the expected projection are due to nonuniformities but do not appear to significantly alter the blow-out regime. For comparison Fig.~\ref{fig:tden} (b) displays a typical distribution measured at the same location when the bunch is photoemitted using the 3-ps long Nd:YLF laser. Varying the charge density was also found to significantly spoil the parabolic profiles. 

\subsection{Longitudinal phase space chirp}
A second set of experiments consisted in measuring the incoming LPS chirp.  Given the initial LPS coordinates of an electron $(z_0, \delta_0)$, the relative momentum offset downstream of the booster cavity operated with an off-crest phase $\varphi$ and accelerating voltage $V$ is 
\begin{eqnarray}
\delta=\frac{\bar{E_0}}{\bar{E}}  \delta_0 - \frac{eVk \sin \varphi}{\bar{E}} z_0, 
\end{eqnarray}
where  $\bar{E}\equiv \bar{E_0}+eV\cos\varphi$ with $\bar{E}_0$ and $e$ being respectively the initial beam's mean energy and electronic charge and $k\equiv 2\pi/\lambda$ (where $\lambda\simeq 0.23$~cm is the wavelength associated to the fundamental mode of the booster cavity). The latter equation is valid provided   $k \sigma_{z,0} \ll 1 $ where $\sigma_{z,0}$ is the incoming bunch length.

The final LPS correlation downstream of the booster cavity is 
\begin{eqnarray}
\mean{z\delta} &=& \mean{z_0\delta_0} \frac{\bar{E_0}}{\bar{E}} -\frac{eV}{\bar{E}} k \mean{z_0^2} \sin \varphi,
\end{eqnarray}
where $\mean{z_0\delta_0}$ is the initial LPS correlation and $\mean{z_0^2} \equiv \sigma_{z,0}^2$ is the squared rms bunch length [the $\mean{u}$ notation indicates the statistical averaging of the variable $u$ over the LPS distribution]. Setting $\varphi=\varphi_0$, where $\varphi_0$ is the phase resulting in an upright  LPS downstream of the booster cavity ($\mean{z\delta}=0$),  provides the incoming LPS chirp
\begin{eqnarray}
{\cal C}&\equiv & \frac{\mean{z_0\delta_0}}{\mean{z_0^2}} = \frac{eV}{\bar{E}_0} k \sin \varphi_0, 
\end{eqnarray}

\begin{figure}[hhhhh!!!!!!!!!!!!]
\centering
\includegraphics[width=0.48\textwidth]{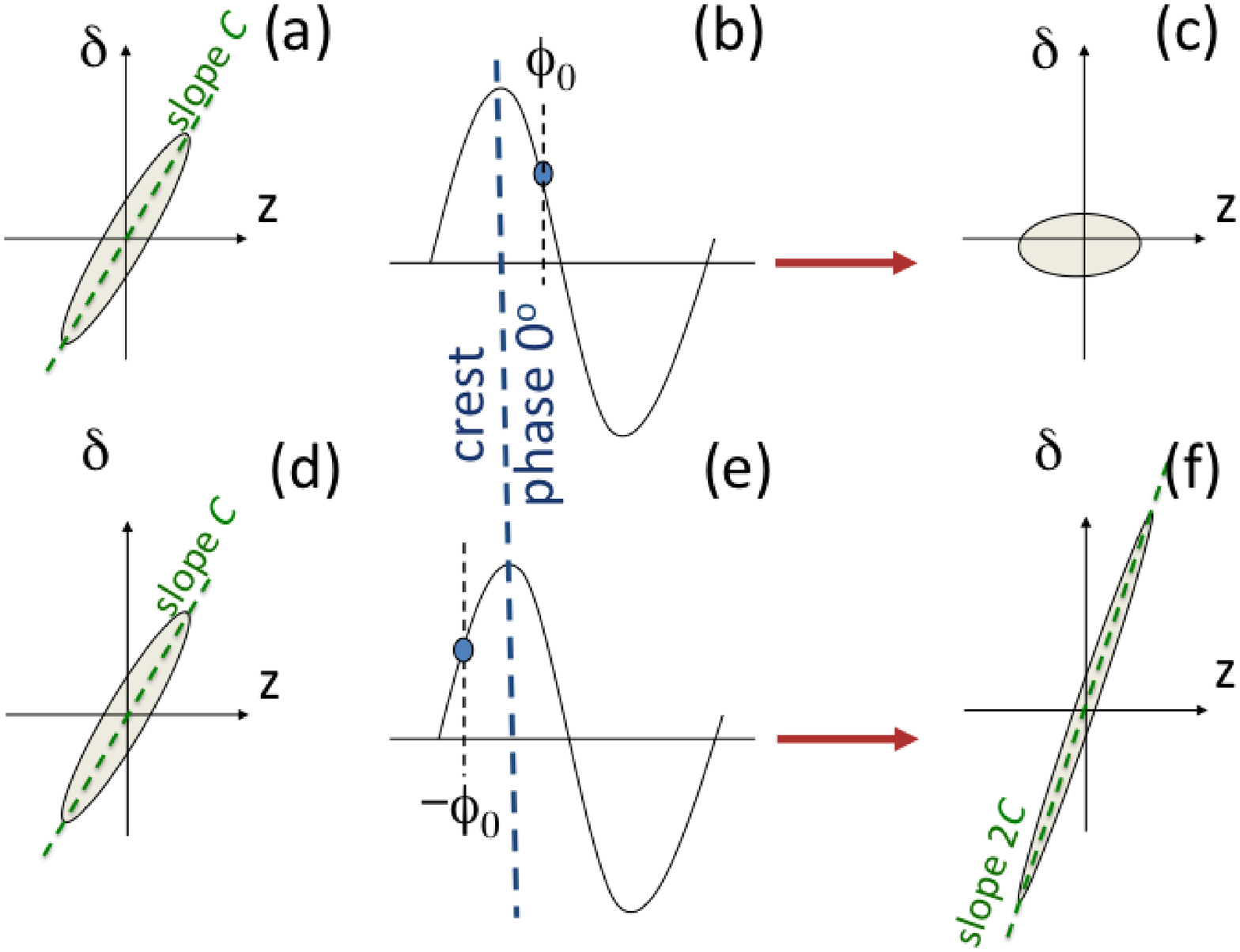}
\includegraphics[width=0.485\textwidth]{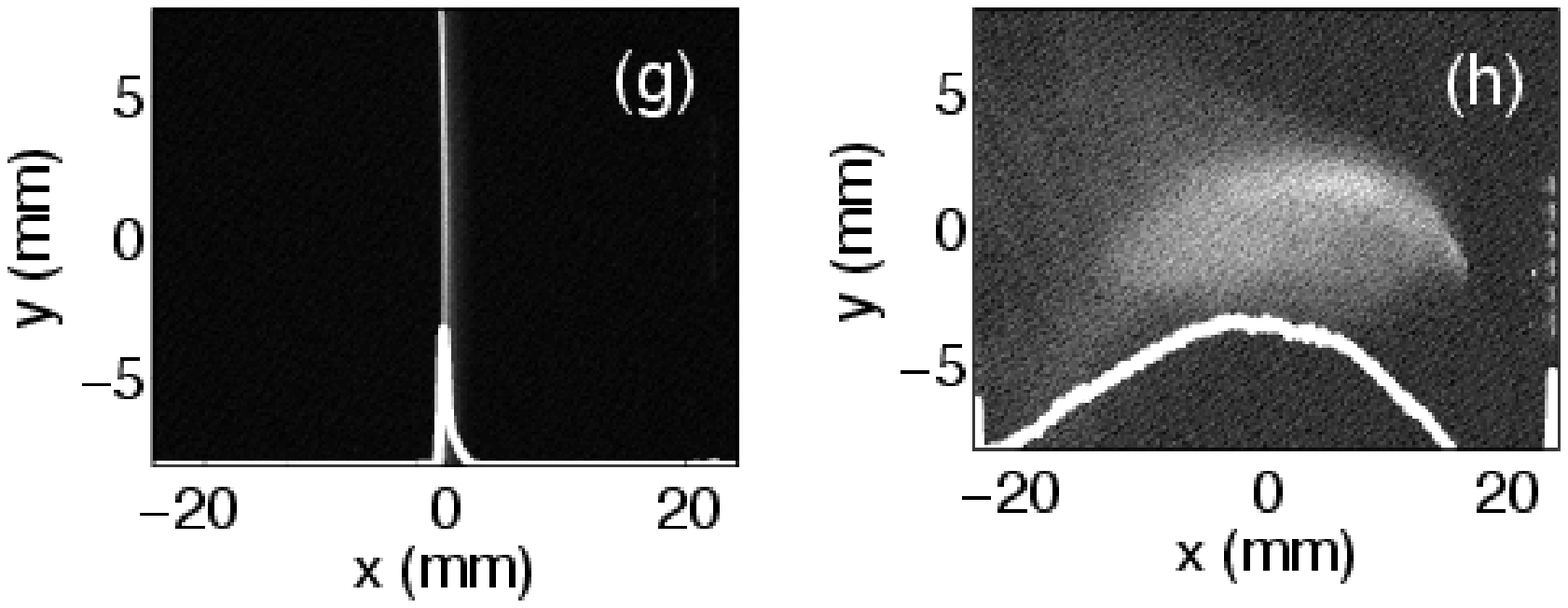}
\caption{Principle of measurement of the LPS chirp [(a), (b), and (c)] and bunch length [(d), (e), and (f)] measurements. The top [plots (a), (d)], middle [plots (b), (e)] and right [plots (c), (f)] columns show respectively  the initial LPS, the location of the bunch (shown as a blue circle) w.r.t. the accelerating voltage, and the final LPS. The images (g) and (h) respectively show an example distribution measured at XS3 location for minimum-energy-spread and off-crest phase settings; see text for details. In these latter images the horizontal axis corresponds to the energy-dispersed direction [$x \mbox{[mm]}\simeq 317 \delta$ where $\delta$ is the fractional momentum spread]. }\label{fig:measscheme}
\end{figure}

\begin{figure}[hhhhh!!!!!!!!!!!!]
\centering
\includegraphics[width=0.45\textwidth]{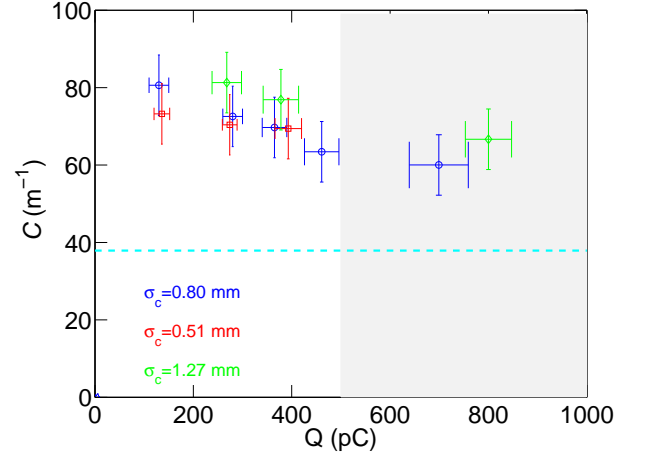}
\caption{Measured longitudinal-phase-space chirp ${\cal C}\equiv \mean{z_0\delta_0}/\mean{z_0^2}$ as a function of charge for three cases of laser transverse rms size $\sigma_c$ on the photocathode surface. The horizontal dashed line represents the nominal chirp when the 3-ps-long Nd:YLF laser is used for photoemission. The shaded area represents the domain where the parabolic character of the temporal distribution was observed to be distorted and is consequently associated to a regime where  a uniformly-filled-ellipsoid bunch is not realized. }\label{fig:chirp}
\end{figure}

Experimentally, $\varphi_0$ can be inferred by minimizing the fractional momentum spread as measured at XS3 as illustrated in Fig.~\ref{fig:measscheme} plots (a), (b, and (c).  In principle the technique could also allow for the measurement of the uncorrelated momentum spread. Unfortunately the resolution of our imaging system was not adequate to support such a measurement: as show in Fig.~\ref{fig:measscheme} image (g), the horizontal width of the distribution is extremely small at the limit of our resolution. Given the operating parameters of the rf gun and booster cavity (respectively $\bar{E}_0$ and $V$), the estimated values of ${\cal C}$ for different charge density are summarized in Fig.~\ref{fig:chirp}.  As discussed in Ref.~\cite{chirp} and observed in Ref.~\cite{moody},  the chirp is not strongly dependent on the charge in the blow-out regime. 

As expected for the blow-out regime, the typical off-crest phase required to minimize the final momentum spread is 
$\sim 55^{\circ}$ which is significantly larger that the phase required for  the longer pulses produced with the nominal Nd:YLF laser ($\sim 25^{\circ}$). Such large off-crest phase is indicative of the strongly correlated longitudinal phase space.

\subsection{Spatio-temporal and current distributions}

In order to measure the spatio-temporal distribution and confirm that the associated current profile is described by a parabolic function we use a variant of the zero-phasing technique~\cite{wang,shaftan}.  The method consists in imparting a chirp on the bunch by operating the booster cavity off-crest.  The beam is then horizontally-dispersed in a  spectrometer. The final horizontal position of an electron at the screen XS3 can be related to the $(x,x',\delta)$ coordinates upstream of the booster cavity as
\begin{eqnarray} 
x_f&=& R_{11}x + R_{12}x' + \eta \delta,
\end{eqnarray}
where the $R_{ij}$'s stand for the transfer matrix elements from the booster cavity exit to XS3 and $\eta$ is the value of the dispersion function at the location of XS3. The fractional momentum offset can be explicited as function of the longitudinal phase space coordinate $(z_0,\delta_0)$ upstream of the booster cavity as 
\begin{eqnarray} \label{eq:xs3} 
x_f &=& R_{11}x + R_{12}x' + \eta \left(\frac{\bar{E_0}}{\bar{E}}  \delta_0 - \frac{eVk \sin \varphi}{\bar{E}} z_0 \right).
\end{eqnarray}
Further writing the initial fractional momentum spread as $ \delta_0 =  \tilde{\delta}_0 + {\cal C} z_0$ where $\tilde{\delta}_0 \ll {\cal C} z_0$ is the uncorrelated contribution and taking $\varphi=-\varphi_0$ in Eq.~\ref{eq:xs3} yields
\begin{eqnarray} \label{eq:xsb} 
x_f&=& R_{11}x + R_{12}x' + \eta \tilde{\delta}_0 + 2 \eta {\cal C} z_0.
\end{eqnarray}

\begin{figure}[hhhhh!!!!!!!!!!!!]
\centering
\includegraphics[width=0.48\textwidth]{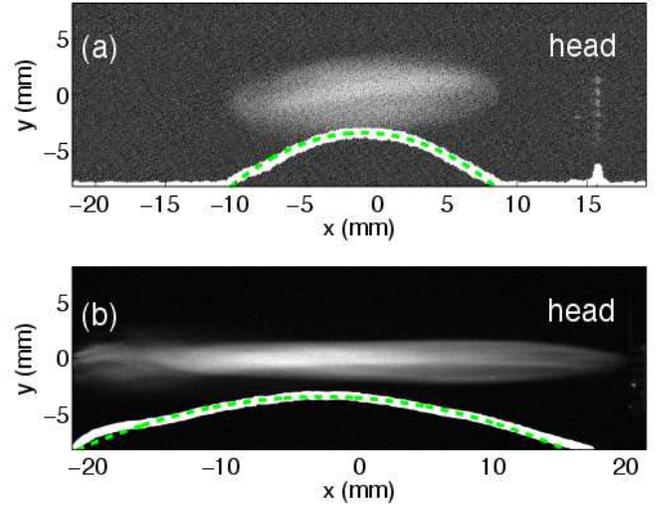}
\caption{Measured spatio-temporal $(x\propto t,y)$ distribution and associated projections (white solid lines) with parabolic fit (green dash lines). The beam density is measured at the Ce:YAG screen XS3 (see Fig.~\ref{fig:beamline}) for $Q=100 \pm 20$~pC (a) and $Q=480~\pm 30$~pC. }\label{fig:ty}
\end{figure}

By properly tuning the quadrupole magnets Q1 and Q2, the horizontal position at XS3 is dominated by the last term in Eq.~\ref{eq:xsb}, i.e., $x_f \simeq 2 \eta {\cal C} z_0$ (see illustration in Fig.~\ref{fig:measscheme}) and the transverse $(x_f,y_f)$ distribution is representative of the $[z_0,y_f(y_0)]$ spatio-temporal distribution. A beam-based calibration procedure was used to infer correspondence to the temporal coordinate by varying the booster cavity phase and recording the beam's centroid motion at XS3. The typical calibration was $24$~pixels/ps. An example of measured spatio-temporal distribution appears in Fig.~\ref{fig:ty}. At low charges the current profiles follow the expected parabolic distribution while the higher charge case starts deviating from the parabolic distribution due to distortion appearing in the tail region. This distortion stems from the image charge becoming more important as the charge increases as previously discussed. %
\begin{figure}[hhhhh!!!!!!!!!!!!]
\centering
\includegraphics[width=0.45\textwidth]{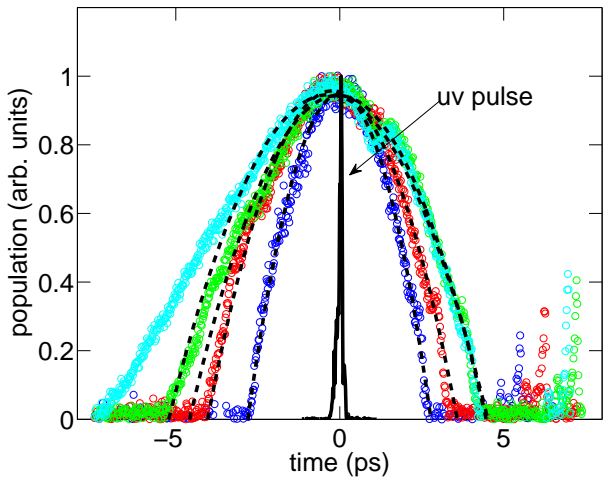}
\caption{Measured peak-normalized  temporal profiles for $Q\simeq130$,~280, 460, and~700~pC (respectively shown as blue, red, green, and cyan circles),  compared with the measured uv laser temporal profile (black solid trace). The dashed lines represent the results of parabolic fits to the measured profiles. The head of the bunch corresponds to $t>0$. }\label{fig:prof}
\end{figure}

\begin{figure}[hhhhh!!!!!!!!!!!!]
\centering
\includegraphics[width=0.45\textwidth]{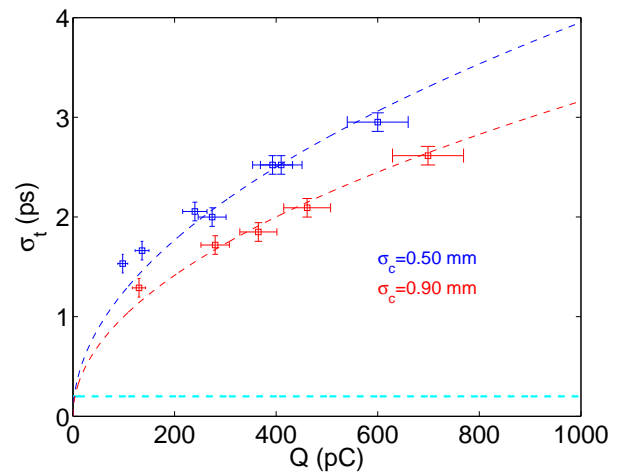}
\caption{Measured rms electron bunch duration (symbols) as a function of charge for two cases of laser transverse spot size $\sigma_c$ on the photocathode. The dashed cyan line represent the uv laser pulse duration. }\label{fig:sigmat}
\end{figure}

Depending on the laser initial spot size, the measured temporal projection follows a parabolic distribution (thereby confirming the ellipsoidal character of the bunch distribution) for charge up to 500~pC ; see Fig.~\ref{fig:prof}.   For the lowest measured charge ($Q=130$~pC), the measured bunch full-width duration is $\sim 7$~ps [corresponding to $\sim 7/(2\sqrt{5}) \simeq 1.6 $~ps (rms)]. This final bunch duration represents a $\sim 10$-fold increase compared to the initial laser-pulse rms duration. Larger expansion ratios are observed for higher-charge bunches. Such a large expansion is a salient feature of the blow-out regime. The rms electron-bunch durations measured as a function of charge for several initial laser transverse size on the photocathode are reported in Fig.~\ref{fig:sigmat}.  As expected for a space-charge-driven expansion, the bunch duration increases for a smaller initial spot size due to the higher initial charge density.

\subsection{Transverse emittance}
Finally, the transverse emittances of the produced bunches were measured for different laser transverse spot sizes $\sigma_c$ and charges $Q$. The best emittance measured was $1.7\pm 0.2$~$\mu$m obtained for the minimum charge $Q=200\pm 20$~pC and $\sigma_c=1.2\pm 0.1$~mm. Although this value is well above state-of-the-art achievements,  it represents a $\sim$ 50\%  improvement compared to the typical emittances measured for similar charge at the A0 photoinjector~\cite{prleex}. 


As pointed in Ref.~\cite{yuelin1}, the blow-out regime does not a fortiori provide the best transverse emittance. In fact a simple argument based on the Eq.~\ref{eq:blowout} and assuming a perfect emittance compensation process indicates that the best achievable transverse normalized emittance scales approximately as 
\begin{eqnarray}
\varepsilon_{\mbox{min}} &=&  \left[\frac{Q}{\pi\epsilon_0 E_0}\frac{kT}{12mc^2} \right]^{1/2} , 
\end{eqnarray}
where $kT$ is the thermal energy and $mc^2$ the electronic rest energy. The latter equation hints that the blow-out regime is generally unfavorable to emittance when operating with high-charge bunches  produced in low-peak-field rf guns. For this reason, directly shaping the laser profile to follow an ellipsoidal distribution would be more effective when an improvement of the transverse emittance is desired~\cite{yuelin2,yuelin3} and for our range of bunch charges. 

\section{Summary}
We have experimentally demonstrated the production of uniformly-filled ellipsoidal bunch in a L-band rf photoinjector by illuminating a Cs$_2$Te photocathode with an ultrashort UV pulse. The bunch ellipsoidal character was shown to be preserved after acceleration to $\sim 14$~MeV with charges up to $\sim 0.5$~nC. The presented work is a significant improvement over a previous attempt to produce ellipsoidal bunches from a Cs$_2$Te photocathode in a L-band rf gun~\cite{oshea}. It also complements the low-charge results obtained for the case of metallic photocathodes in Ref.~\cite{pietro} 

Further studies are planned at the High-Brightness Electron-beam Source Laboratory (HBESL) which will be commissioned following the
A0PI decommissioning. At HBESL, a 3.9-GHz deflecting cavity located downstream of the rf gun will enable a precise characterization of the longitudinal phase space and  possibly the exploration of the response time of Cs$_2$Te photocathodes. 

Finally,  the results presented in this paper support the possible implementation of the blow-out regime at the Advanced Superconducting Test Accelerator (ASTA), currently in construction at Fermilab~\cite{church}, which will incorporate a Cs$_2$Te photocathode. Such an implementation at ASTA would enable the production of  high-peak-current (10's~kA) electron bunches to be used toward advanced accelerator R\&D.  

\section{Acknowledgments}
We are indebted to R. Montiel, J. Santucci, and B. Tennis for their excellent operational and technical supports.  We thank M. Church,
H. Edwards, E. Harms, A. H. Lumpkin, and V. Shiltsev for their interest and encouragement. One of us (PP) wishes to thank Dr. Smith of
AS-Photonics and Dr. Gilevich of SLAC for their help with {\sc snlo}. This work was supported by DOE award DE-FG02-08ER41532 and DOD DURIP award N00014-08-1-1064 to Northern Illinois University and by the DOE contract DE-AC02-07CH11359 to the Fermi Research Alliance LLC.


\begin{thebibliography}{99}
\bibitem{comment} The recent study reported in Ref.~\cite{oshea} does not use a short laser pulse and cannot explore the possible limitations coming from the Cs$_2$Te response time. In addition the reported experimental results (e.g.  Fig.~6 of Ref.~\cite{oshea}) do not provide any evidence that the conditions for the blow-out regime were actually attained: as noted by the authors, the image charge strongly affected the ellipsoidal self-shaping process.
\end{thebibliography}
\end{document}